\documentclass[article,11pt]{revtex4}

\usepackage{graphicx}
\usepackage{epsfig} 
\usepackage{epstopdf}
\usepackage{xcolor}
\usepackage{amsfonts}
\usepackage{amssymb,amsmath}
\usepackage{appendix}
\usepackage{setspace}
\usepackage{lipsum}
\usepackage{enumerate}
\usepackage{hyperref}

\usepackage{soul} %Strikethrough text

\begin{document}

\title{Controlling the average degree in random power-law networks}

\author{Allan Vieira, Judson Moura, Celia Anteneodo\\ 
 Department of Physics, PUC-Rio, Rua Marqu\^es de S\~ao Vicente, 225, 22451-900, Rio de Janeiro, Brazil }

%\date{August 2021}

\begin{abstract}
    We describe a procedure that allows continuously tuning the average degree $\langle k \rangle$ of uncorrelated networks with power-law  degree distribution $p(k)$. Inn order to do this, we modify  the low-$k$ region of $p(k)$, while preserving the large-$k$ tail up to a cutoff. Then, we use the modified $p(k)$ to obtain the degree sequence required to construct networks  through the configuration model.  We analyze the resulting nearest-neighbor degree and local clustering to verify the absence of  $k$-dependencies. Finally,  a further modification is introduced to eliminate the sample fluctuations in the average degree.
\end{abstract}

\maketitle

\section{Introduction}

Artificial networks are an important substrate for studying the dynamics of many complex systems. In particular scale-free, or more generally power-law networks, have been widely used for that purpose. 
The power-law exponent that characterizes the decay of the degree distribution is a crucial quantity that can produce drastic changes in the phenomenology of the system. But the average degree can also play an important role as a control parameter promoting critical phenomena (see, for instance, ~\cite{watts,extremists,qvoter,silvio,cooperation}).  Moreover, it can influence other structural measures of a network, such as average nearest neighbor degree ~\cite{brito_correlations}. 
Hence, the average degree needs to be taken into account for unbiased comparisons~\cite{brain2010}. 
 
However, it is an often neglected quantity when building networks.
As we will see below, in this section,  the resulting average connectivity of a network may present significant  deviations from the initially proposed value, especially for power-law networks due to their heterogeneity.
Then, our purpose is to present a simple procedure that allows to adjust the average degree in random networks that are constructed via the configuration model~\cite{CM,newman}.

Let us consider distributions of degrees with the power-law  form
\begin{equation}
    p(k)= \frac{\cal N}{k^\gamma},  \hspace{1cm} \mbox{for $k_{min}\le k\le k_{max}$},
    \label{eq:pk}
\end{equation}
with $\gamma >2$ and where ${\cal N}=\sum_{k=k_{min}}^{k_{max}} k^{-\gamma}$ is the normalization constant. 
We will consider that the minimal degree can take values $ k_{min}\ge 2$ (hence, excluding only nodes with one link), and 
$k_{max}$ is a maximal allowed value (cutoff). 
The natural cutoff is $k_{max} \propto N^{1/(\gamma-1)}$~\cite{mendes2002}, but  
a structural cutoff  $k_{max}\propto \sqrt{N}$ has been considered in the literature~\cite{UCM2004, UCM} to reduce  $k$-dependencies in such networks.

In order to build a network, after randomly drawing the degree sequence  from the distribution (\ref{eq:pk}), 
we link the nodes according to the configuration model~\cite{CM,newman}. 
When drawing the degrees for each network, the effective maximal degree will be $k^*_{max}\le k_{max}$ and the mean degree of the network, $\bar{k}$, will in general differ from the average computed with the distribution  (\ref{eq:pk}) 
\begin{equation} \label{eq:kk}
    \langle k\rangle= \sum_{k=k_{min}}^{k_{max}} k\, p(k)\,.
\end{equation}
%, also  checked for disconnected components. 

\begin{figure}[h!]
    \centering
  \includegraphics[scale=0.9]{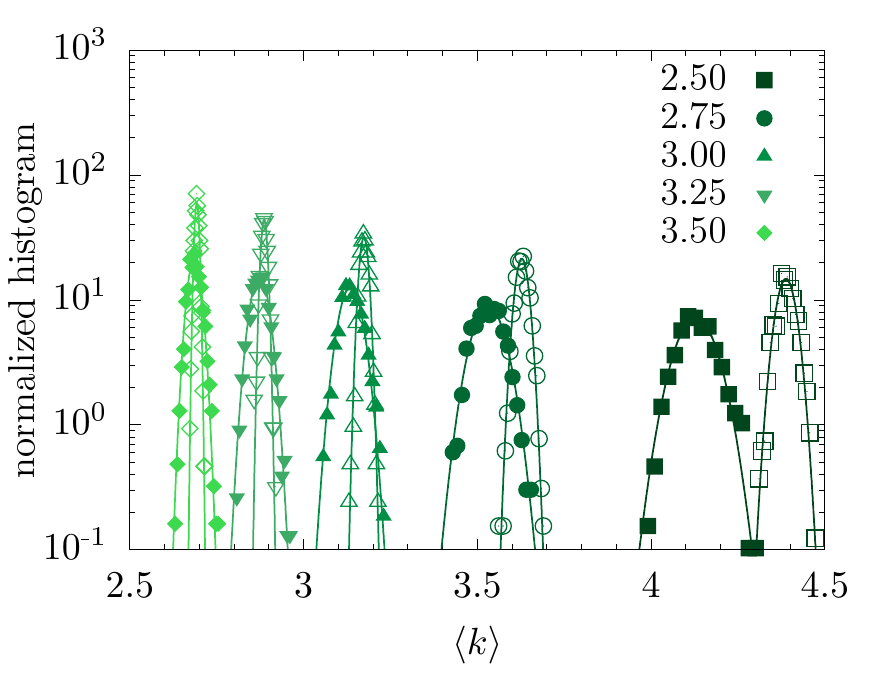}
    \caption{Normalized histogram of the average connectivity $\bar{k}$ of the samples, computed over 1000 realizations drawn from $p(k)$, for different values of the exponent $\gamma$ given in the legend, 
    and two values of the network size $N$: $10^4$ (filled) and $10^5$ (hollow symbols). We used $k_{min}=2$ and the   prescription $k_{max}=\sqrt{N}$~\cite{UCM}. 
    The solid lines are Gaussian curves centered in $\langle k \rangle$, fitting the  width, which decays with the network size as $1/\sqrt{N}$. 
    }
    \label{fig:average}
\end{figure}

An illustration is given in Fig.~\ref{fig:average}, for networks with  different values of $\gamma$, and two different sizes $N$. 
In all cases
$k_{min}=2$ and the   prescription $k_{max}=\sqrt{N}$ was used. 
For each sample network, $\bar{k}$ was recorded, and the histogram of values over 1000 realizations is shown in Fig.~\ref{fig:average}.
The deviation of $\bar{k}$ around $\langle k \rangle$  decreases with $N$, as  $1/\sqrt{N}$, as expected. But, when varying $\gamma$,  a large spread of values of $\bar{k}$ emerges. This spread increases  with $N$, although the deviation for each value of $\gamma$ decreases with $N$. 
Moreover, recall that in the infinite network limit, the moments of order $n\ge \gamma-1$ are divergent.

\begin{figure}[h!]
    \centering
\includegraphics[scale=0.45]{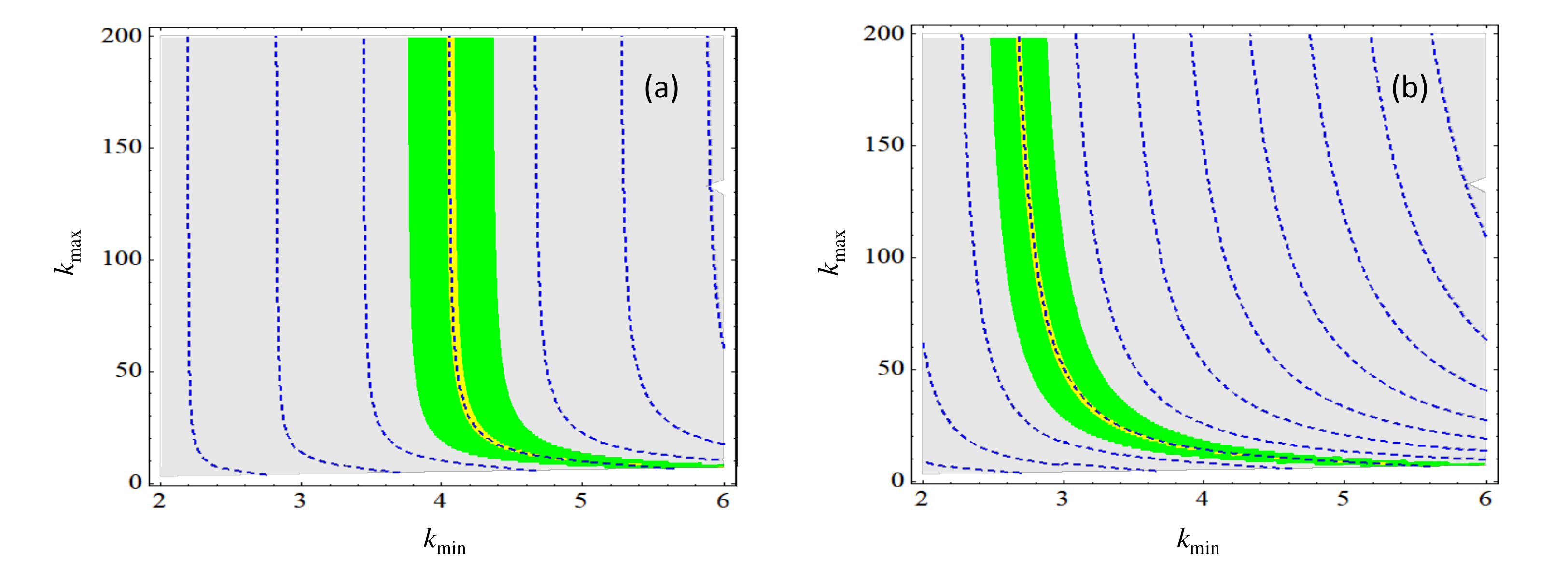}
    \caption{Regions in the plane $k_{min}-k_{max}$ for which the average connectivity $\langle k\rangle$ is within the intervals $k_{max}>k_{min}$ (lightgray)
     [5.5,6.5] (green) and [5.9,6.1] (yellow), for  power-law networks with exponent
    $\gamma=3.5$ (a) and 2.5 (b). 
   The dashed curves correspond to $\langle k\rangle=3,4,\dots$.  
    Despite the continuous representation, recall that only integer values of $k_{min}$ and $k_{max}$ are allowed, which restricts even more the realization of a chosen value of $\langle k \rangle$.
    }
    \label{fig:kminkmax}
\end{figure}

Furthermore,
if we try to fix $\langle k\rangle$,   restrictions emerge.
For instance, for $\gamma=2.5$, and choosing  
 $k_{max} = \sqrt{N}$ ( $\approx 316$, when  $N=10^5$), 
 and $k_{min}=2$ (or alternatively 3), 
 then, the average degree is   $  \langle k \rangle \simeq 4.39$ (or alternatively $\simeq 6.97$). This means that increasing $k_{min}$ in one unit  produces an increase of about 2.6 in the average degree. 
More generally, 
for given $\left( k_{min} , k_{max} , \gamma \right)$, 
the outcome of a given value of $\langle k\rangle$, within an allowed  tolerance interval,  is restricted 
to a very narrow region in the plane  $k_{min}-k_{max}$, as can be observed in Fig.~\ref{fig:kminkmax}, for two different values of $\gamma$. Moreover,  since  $k_{min}$ is integer, 
combinations with a low tolerance may be not feasible. 
Also notice that the average degree becomes insensitive to large enough $k_{max}$, while the high probability of the small degrees turns $\langle k \rangle$ very sensitive to $k_{min}$. 

In this context, our goal is to modify the degree distribution to adjust
$\left \langle k \right \rangle$, given the set of parameters $\left( k_{min} , k_{max} , \gamma \right)$. 
In particular, we will take into consideration that, as we have seen in Fig.~\ref{fig:kminkmax}, a limitation comes from the integer character of $k_{min}$, and also from the high probability that the smallest values of $k$ have in a pure power-law distribution.
Therefore, the schemes that we will propose somehow emulate an analytic continuation between consecutive integer values of $k_{min}$, thus allowing to obtain a continuum of values of $\langle k \rangle$, overcoming the limitations shown in Fig. Fig.~\ref{fig:kminkmax}.

The proposed schemes are defined in Section~\ref{sec:teo}. 
In Section~\ref{sec:correl}, we check that the adjustments of 
$p(k)$ do not introduce undesired correlations in the networks constructed via the configuration model. In Section~\ref{sec:zero}, we describe a procedure to eliminate the fluctuations in the average degree for finite networks.
Final considerations are presented in Sec.~\ref{sec:final}.

\section{Modified degree distribution}
\label{sec:teo}

\begin{figure}[h!]
    \centering
    \includegraphics[scale=0.8]{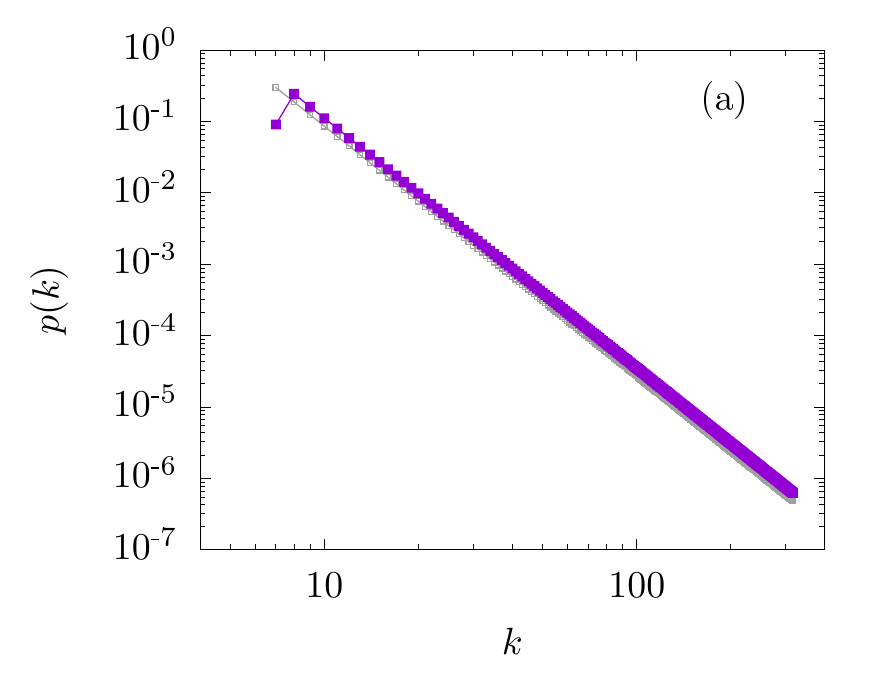}
     \includegraphics[scale=0.8]{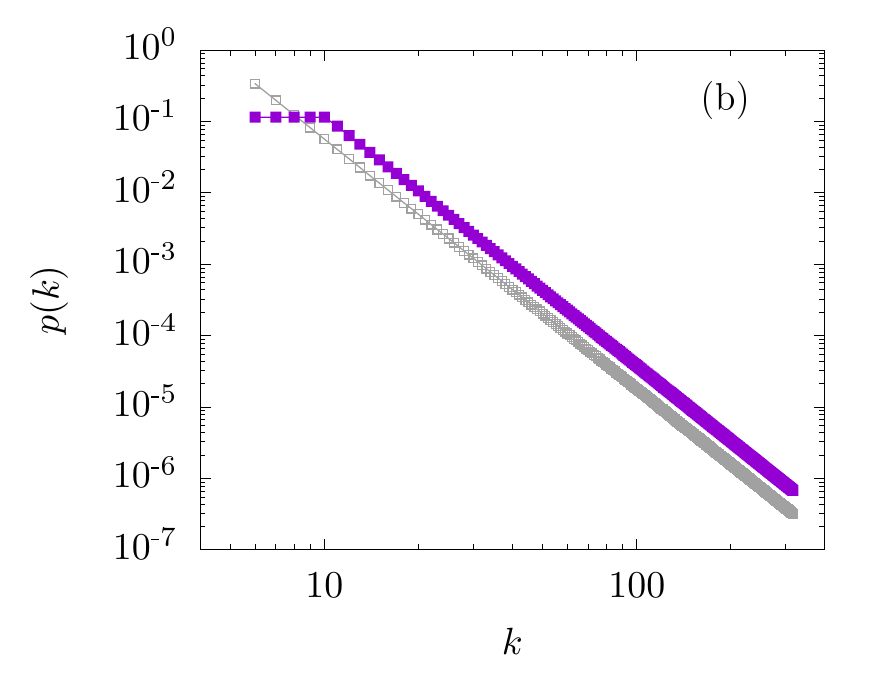}
    \caption{Schematic representation of schemes I (a) and II (b), in lilac filled symbols. 
    In both cases,   $\gamma=3.5$,  $N=10^5$, $  \langle k \rangle=12$, $k_{max}=\sqrt{N} \simeq 316$. 
  The corresponding pure power-law $p(k)$ is represented by light-gray hollow symbols.
    }
    \label{fig:theory}
\end{figure}

Based on the introductory discussion, we propose two simple schemes to adjust the average degree of a power-law network, 
with $p(k)$ given by Eq.~(\ref{eq:pk}). These schemes, illustrated in each panel of Fig.~\ref{fig:theory},  consist of modifying the low-degree region of the original  $p(k)$. 
In case I, only the minimal degree does not follow a power-law. In case II, several points can form a plateau, before the power-law decay.  
Notice that, by altering the shape of low-$k$ region, related to the more likely values of $k$, concomitantly the probability associated with the tail increases, allowing for more highly connected nodes.
As we will see, this slight modification allows to adjust the average degree. 

 It is also noteworthy that real degree distributions do not follow a pure power-law but are better represented by the distributions that result from the modified schemes. 
 For example,  networks of word co-occurrence, company directors and internet autonomous systems resemble the form of  scheme I, while (scientists, film actors) collaboration networks  resemble the shape of scheme II~\cite{newman-chapter}.

\subsection{Scheme I}

\begin{figure}[h!]
    \centering
    \includegraphics[scale=0.4]{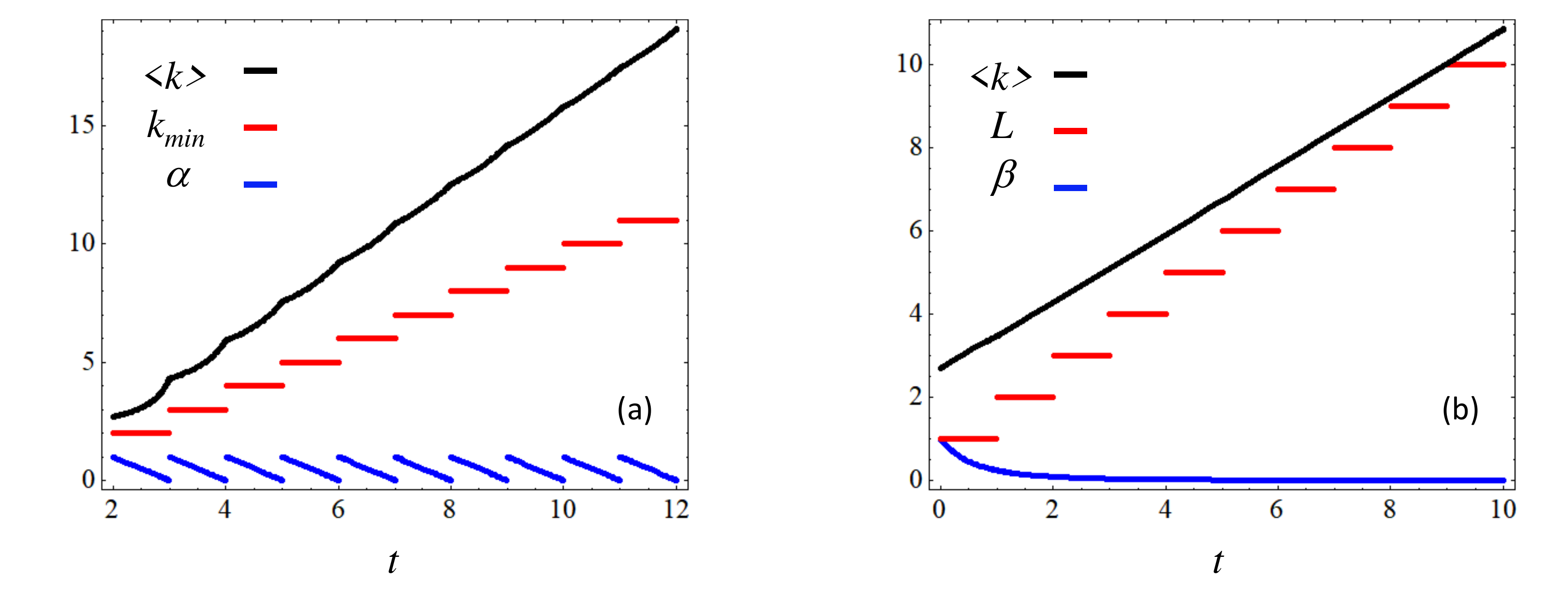}
    \caption{Parameters of the degree distribution vs. the control parameter $t$: (a) $k_{min}$ and $\alpha$  in Eq.~(\ref{eq:p1}) for scheme I
    (b) $L$ and $\beta$, with $k_{min}=2$  in Eq.~(\ref{eq:p2}) for scheme II. The corresponding value of $\langle k \rangle$ is also shown. }
    \label{fig:pars}
\end{figure}

 Starting from a pure power-law distribution, we scale the probability of the minimal degree $k_{min}$ by a factor $\alpha\in (0,1]$ and renormalize the whole distribution, 
as illustrated in Fig.~\ref{fig:theory}(a).  
That is 
\begin{equation}
    p(k)=\left\{
    \begin{array}{ll}
     \alpha\, k_{min}^{-\gamma}/Z,     & \mbox{if $k=k_{min}$}, \\[3mm]
       k^{-\gamma}/Z,    &  \mbox{if $ k_{min}<k\le k_{max}$},
    \end{array}  \right. 
\end{equation}
where  $Z=  \alpha\,  k_{min}^{ -\gamma}+\sum_{k > k_{min}}^{k_{max}} k^{-\gamma}$ is the normalization constant.  
In order to do that, let us consider  the parametrization 
\begin{eqnarray} \nonumber
k_{min}(t) &=&  \lfloor t \rfloor, \\ \label{eq:p1}
\alpha(t) &=&  \lfloor t \rfloor-t+1\,,
\end{eqnarray}
where $t\ge 2$ and  $\lfloor \cdots \rfloor$ is the floor function. 
In the particular case $t=2$, we have  $k_{min}=2$ and $\alpha=1$. Increasing $t$ between consecutive integer values, 
makes $\alpha$ decrease from 1 to 0. When  $t$ reaches an integer value,  $\alpha$ is reset to unit, while $k_{min}$ increases  in one unit. Therefore, when $t$ is integer, the pure power-law distribution starting from a minimal degree $k_{min}=t$ is recovered. For non-integer $t$,   the distribution has the shape illustrated in Fig.~\ref{fig:theory}(a). 
The plots of the coefficients of the degree distribution and of the average degree, versus the control parameter $t$, are shown in Fig.~\ref{fig:pars}(a).
Note that, for given $\gamma$ and $k_{max}$,  
tuning $t$ allows to change continuously the value of the average degree,  
which is given by
\begin{equation}
    \langle k\rangle(t) = \frac{  \alpha \,  k_{min}^{1-\gamma}+\sum_{k=k_{min}+1}^{k_{max}} k^{1-\gamma} }{ \alpha\,  k_{min}^{ -\gamma}+\sum_{k=k_{min}+1}^{k_{max}} k^{-\gamma}}\,,
\end{equation}
where $k_{min}$ and $\alpha$ are given by Eqs.~(\ref{eq:p1}).

\subsection{Scheme II}

Another modification of the pure power-law distribution is illustrated in Fig.~\ref{fig:theory}(b), where a plateau (meaning  uniform distribution for  the $L$ lowest degrees)  is considered. 
The modified distribution is
\begin{equation}
    p(k)=\left\{
    \begin{array}{ll}
     \beta/Z\,,   & \mbox{if $k< k_{min}+L$}, \\[3mm]
        k_{min}^\gamma/ (k^{\gamma} Z)\,,    &  \mbox{otherwise},
    \end{array}  \right. 
\end{equation}
where $Z=\beta L 
     +\sum_{k=k_{min}+L}^{k_{max}} (k_{min}/k)^{\gamma}$ is the normalization constant.

%%% L= kbeta-kmin+1

In this case,  there can be more combinations  to obtain a given average degree, since there is an additional parameter. In fact, by 
choosing $k_{min}$, 
the following parametrization
controls the length and level of the initial plateau: 
\begin{eqnarray} \nonumber
L(t) &=&  \lfloor t \rfloor +1 \equiv\lceil t \rceil , \\ \label{eq:p2}
\beta(t) &=& \left( \frac{k_{min}}{k_{min}+t} \right)^\gamma\,,
\end{eqnarray}
where we can take $t\ge 0$. For $t=0$, 
we recover the pure power-law. 
The behaviors of  $L$ and $\beta$ with $t$, for $k_{min}=2$, are shown  in Fig.~\ref{fig:pars}(b), together with the corresponding average degree  
 \begin{equation}
    \langle k\rangle(t) = \frac{  \beta \sum_{k=k_{min}}^{k_{min}+L-1} k
     +\sum_{k=k_{min}+L}^{k_{max}} k(k_{min}/k)^{\gamma} }{ \beta L
     +\sum_{k=k_{min}+L}^{k_{max}} (k_{min}/k)^{\gamma}}\,, 
\end{equation}
where $L$ and $\beta$ are given by Eqs.~(\ref{eq:p2}).

\section{Correlation analysis}
\label{sec:correl}

Based on the modified degree distributions, we built networks using the configuration model. Self and multiple connections are not frequent in the cases considered and realizations containing such connections were discarded. 
Similarly, only networks with  giant component size $N$ were considered.

In order to identify degree-degree correlations, we consider
the average degree of the neighbors of a given node $i$,  $k_{nn,i}$, which in terms of the adjacency matrix $A$, is given by      $k_{nn,i} = \frac{1}{k_i} \sum_{j} k_jA_{ij}$. 
We compute the average over all nodes with the same degree, $k_i=k$,
\begin{equation}
    k_{nn}(k) = \frac{1}{N_k} \sum_{i|k_i=k} k_{nn,i},
    \label{eq:knn(k)}
\end{equation}
where  $N_k$ represents the number of nodes with degree  $k$. This quantity defines the level of assortativity of the network. If $k_{nn}(k)$ is an increasing (decreasing) function of $k$, the network is  assortative (dissassortative).
For uncorrelated networks, it is $k$-independent, and given by~\cite{newman-chapter,serrano2007correlations}  
\begin{equation}
   k_{nn}^{unc} =  \frac{\langle k^2 \rangle}{\langle k \rangle}.
\label{eq:knn-unc}
\end{equation}

We also consider the local
clustering coefficient $c_i$, given by the ratio of the number of existing connections between neighbors of site $i$, $e_i =  \sum_{j,k} A_{ij}A_{jk}A_{ki}$  (triangles), over its total possible number $ k_i(k_i - 1)/2$. 
 If $k_i=0,1$, then $c_i = 0$. 
 Grouping the local clustering $c_i$ of those vertices with the same degree $k$ (as done in Eq. (\ref{eq:knn(k)})), we have
 \begin{equation}
    c(k) = \frac{1}{N_k} \sum_{i|k_i=k} c_i,
    \label{eq:c(k)}
\end{equation}
 which in 
an uncorrelated network is $k$-independent and given by~\cite{newman}  
\begin{equation}
   c^{unc}  =  \frac{(\langle k^2 \rangle - \langle k \rangle)^2}{N \langle k \rangle^3}\,.
\label{eq:c-unc}
\end{equation}

\begin{figure}[b!]
    \centering
 \includegraphics[scale=0.8]{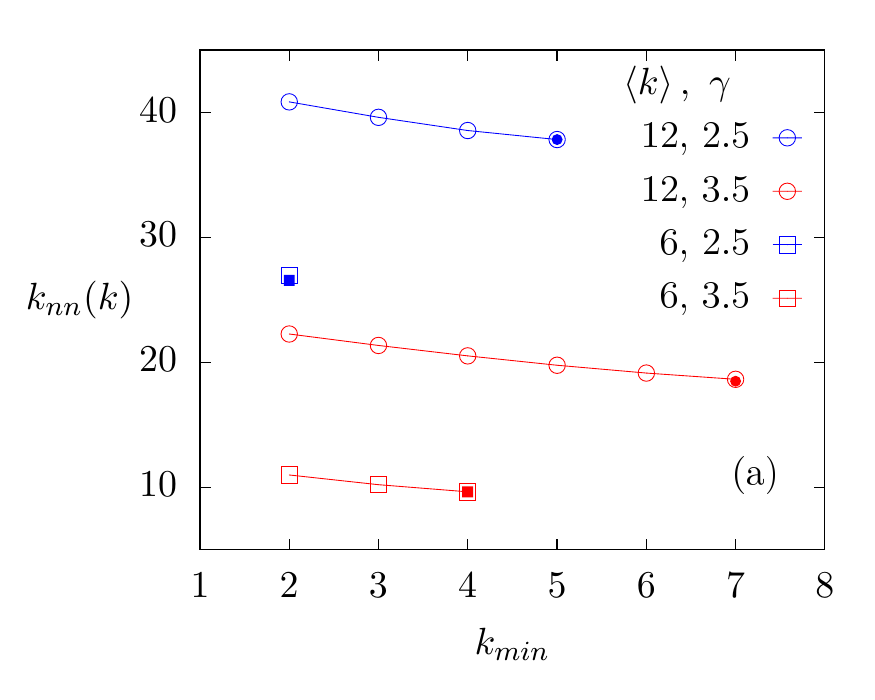}
   \includegraphics[scale=0.8]{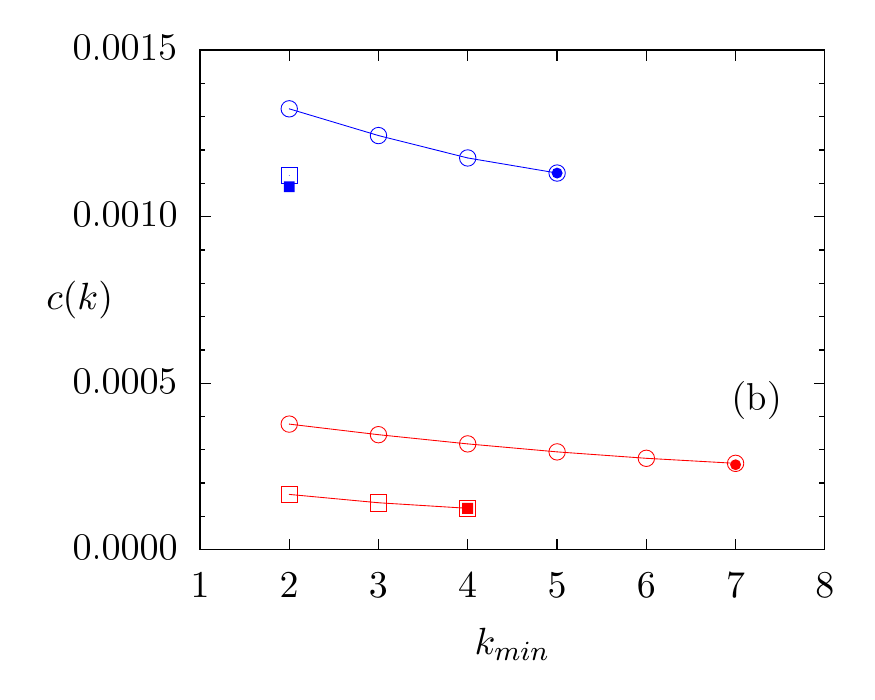}
    \caption{
  Effect of schemes I (full symbols) and II (hollow symbols) on the values of  (a) $k_{nn}^{unc} $ and (b) clustering $c^{unc}$ as a function of $k_{min}$, for two different $\gamma$: 2.5 (blue) and 3.5 (red) and different values of $\langle k \rangle$ (joined by lines). In all cases $k_{max}=\sqrt{N}$, and $N=10^5$.
     }
    \label{fig:teoricos}
\end{figure}

In  Fig.~\ref{fig:teoricos},  we show the theoretical values of  $k_{nn}^{unc}$ and $c^{unc}$ for degree distributions following schemes I (filled symbols) and II (hollow symbols), for two different values of $\langle k \rangle$ and two values of $\gamma$. 
Using scheme II, there can be more choices of the minimal degree $k_{min}$, since one can tune the length of the plateau, which provides and extra parameter. 
Notice that the degree distribution of scheme I yields minimal values of $k_{nn}^{unc}$ and $c^{unc}$, associated to the reduction of the occurrence of low-$k$ nodes.  

\begin{figure}[h!]
    \centering
 \includegraphics[scale=0.8]{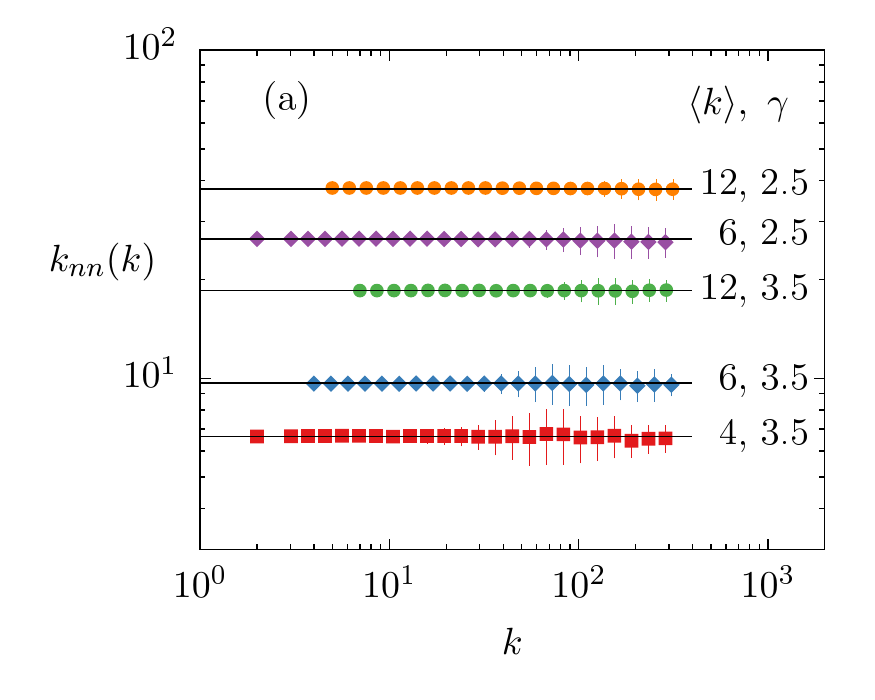}
   \includegraphics[scale=0.8]{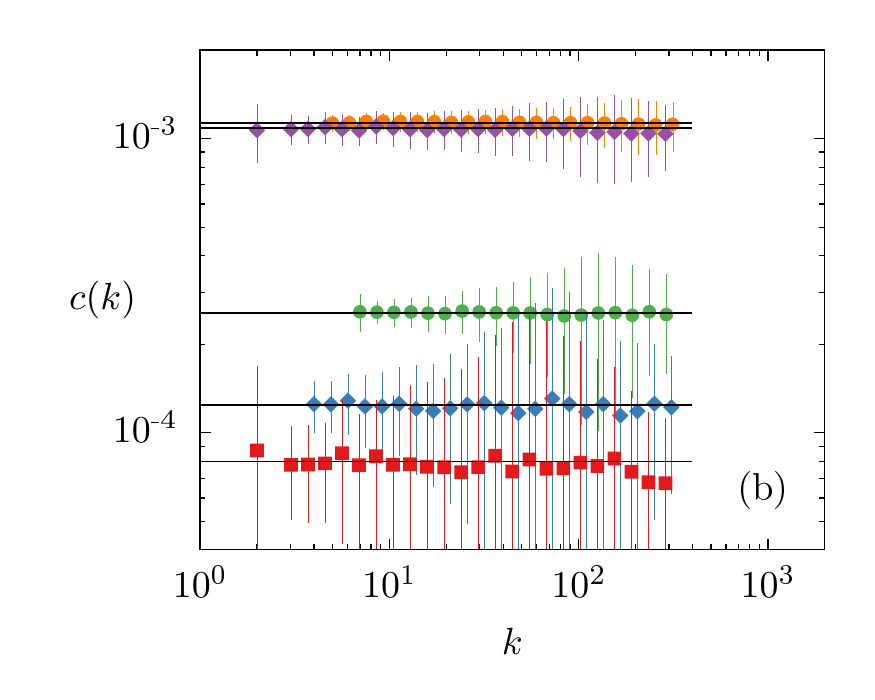}
    \caption{
  (a) $ {k}_{nn}(k)$ and (b) $c(k)$ vs. $k$, for networks constructed using the CM, after applying scheme I, with different values of 
  ($\langle k \rangle,\gamma$) indicated in the figure, and suitable  choices of $k_{min}$ that can be identified by the starting point of each curve. The symbols correspond to the average  and the vertical lines to the standard deviation of the data in each bin, computed over 100 networks.
     The horizontal lines correspond to the uncorrelated values $k_{nn}^{unc}$ and $c^{unc}$ given by Eq.~(\ref{eq:knn-unc}) and Eq.~(\ref{eq:c-unc}), respectively.
    }
    \label{fig:knn_celia}
\end{figure}

For the constructed  networks, we calculated $k_{nn}(k)$, the average degree of the neighbors, and the clustering $c(k)$.
The results for scheme I are presented  in Fig.~\ref{fig:knn_celia}. 
The horizontal  lines correspond to the values predicted for uncorrelated networks.
A very good agreement is observed between measured and theoretical values indicating that correlations are not introduced by the correction scheme, as expected. Same agreement is observed for networks constructed with scheme II (not shown).

\begin{figure}[h!]
    \centering
 \includegraphics[scale=0.8]{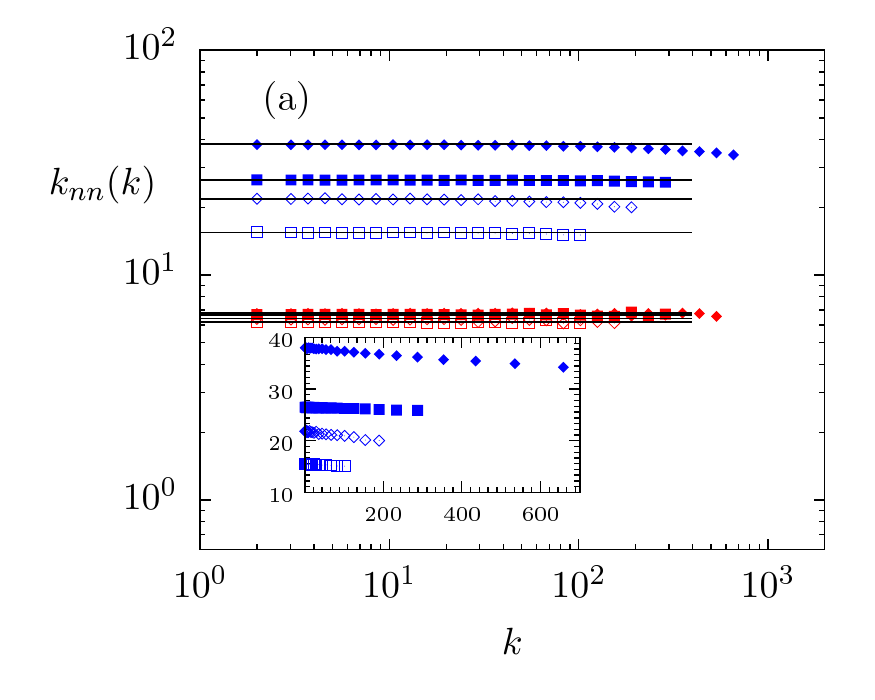}
   \includegraphics[scale=0.8]{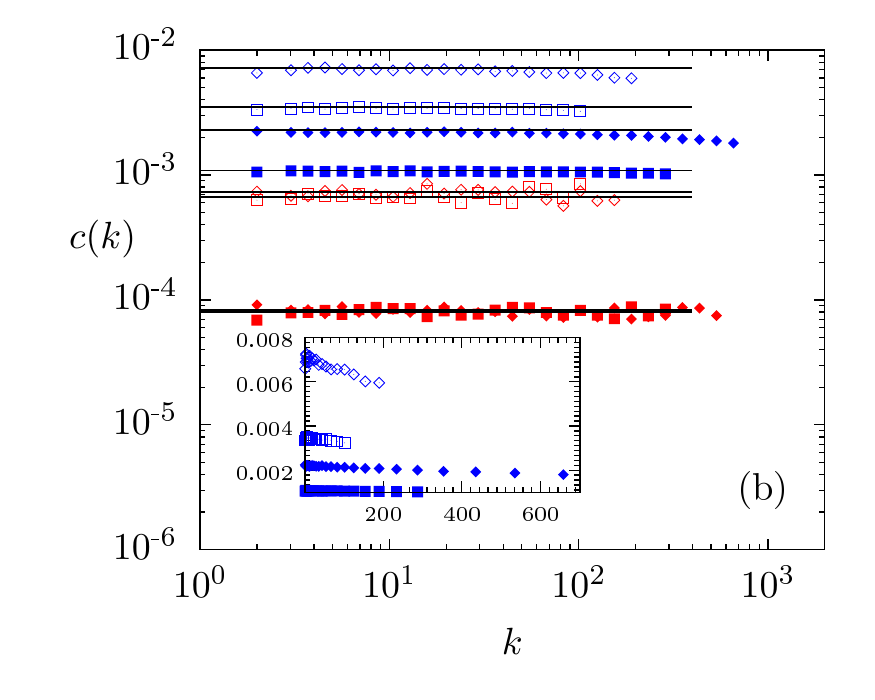}
    \caption{ Size effects. 
  (a) $ {k}_{nn}(k)$ and (b) $c(k)$ vs. $k$, for networks built based on scheme I, with $k_{min}=2$, average degree $\langle k\rangle=6$   and two values of $\gamma$: 2.5 (blue) and 3.5 (red). The symbols correspond 
  to the average over 100 networks (in this case error bars were not plotted for clarity).
  Two different network sizes were considered $N=10^4$ (hollow symbols) and $N=10^5$ (filled symbols). 
  Moreover, two different values of $k_{max}$ were considered: $\sqrt{N}$ (squares) and $\sqrt{\langle k \rangle N}$ (diamonds). 
     The horizontal lines correspond to the uncorrelated values $k_{nn}^{unc}$ and $c^{unc}$ given by Eq.~(\ref{eq:c-unc}) and Eq.~(\ref{eq:knn-unc}), respectively. 
     The insets show the same data for $\gamma=2.5$ of the main frame using linear scales. %Each line is a linear fit.
     }
    \label{fig:N&kmax}
\end{figure}

In Fig.~\ref{fig:N&kmax},  for networks based on scheme I, we show the effect of changing $N$ and the structural cut-off $k_{max}$, for two values of $\gamma$. Notice that larger $N$ produces the increase of $k_{nn}$ and the decrease of $c$. 
For both network sizes the cut-off $k_{max}=\sqrt{N}$ (squares) allows to keep $k_{nn}$ nearly constant for all $k$. 
The cut-off $k_{max}=\sqrt{\langle k\rangle N}$ (diamonds) is enough to avoid $k$ dependencies for $\gamma=3.5$ but not  in the case $\gamma=2.5$ where  the degree distribution has heavy tails. 
In fact, in the latter case, a noticeable decay  of $k_{nn}$ and $c$ with $k$ is observed, although this effect is reduced by increasing system size.  
Actually, this dependency which is nearly linear is always present, but becomes negligible for appropriate $k_{max}$. This effect is not a consequence of the introduced schemes, but it is also observed when the degree distribution is a pure power-law~\cite{UCM}.

\section{Eliminating fluctuations in the average degree}
\label{sec:zero}

The two schemes proposed above fulfill the function of controlling the average degree of power-law networks.
However, as exemplified in figure \ref{fig:average}, the sample average $ \bar{k}$ can still fluctuate around the prefixed value 
$\langle k \rangle$, for finite networks. This is because the process by which the sequence of degrees is constructed is purely random. In this subsection, we present a way to  eliminate the fluctuations in the average degree. 
The procedure consists of two steps to  chose the sequence of degrees. The first step is deterministic and the second stochastic.

To generate the sequence of degrees, we start from a distribution $p^*(k)$. The expected value for the number of  vertices with degree $k$ is $N_k = N p^*(k)$, which in general is not an integer value, as it must be for an individual realization. 
Then, in order to obtain integer number of vertices,  we perform a first deterministic step, where we truncate $N_k$. Then, we add to the list  $\lfloor N p^*(k) \rfloor$  times the degree $k$.
This procedure produces fewer vertices and edges than required, which will be corrected in a second stage. 
 
\begin{figure}[h!]
    \centering
    \includegraphics[scale=0.76]{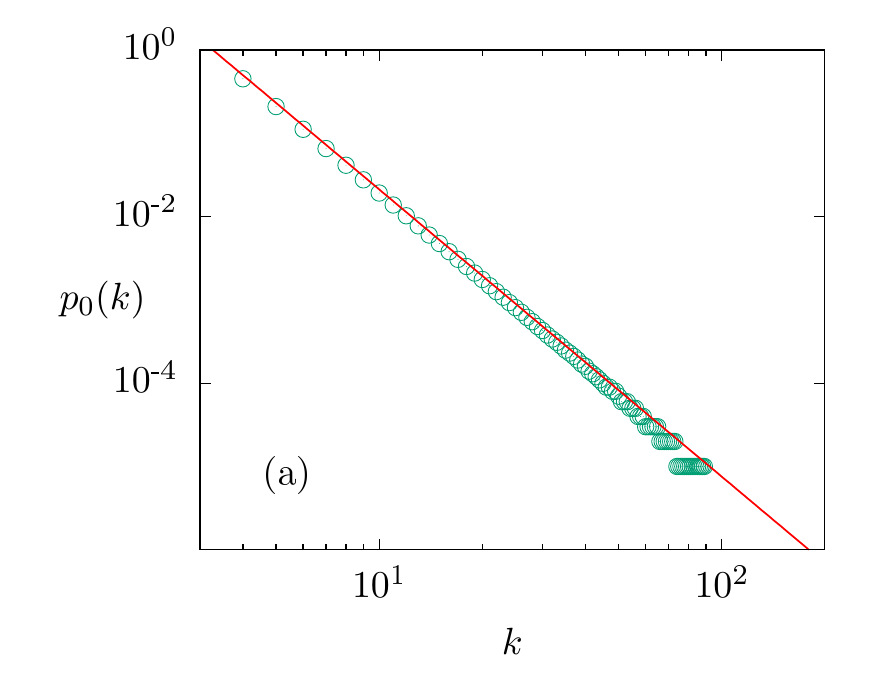}
    \includegraphics[scale=0.76]{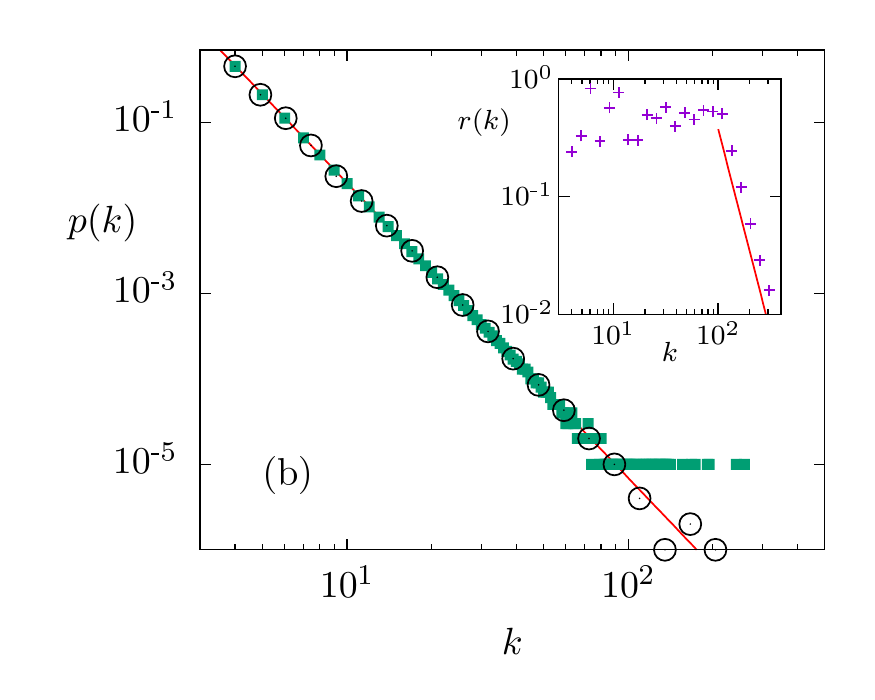}
    \caption{(a) Deterministic non-normalized $p_0(k)=\lfloor N p^*(k) \rfloor/N$ (hollow green symbols), where the original $p^*(k)$ is a pure power-law with exponent $\gamma \sim 3.45$, %$k_{min} = 4$,  $k_{max}\simeq\sqrt{N}$,
The number of vertices was initially $N=10^5$, 
    but after the truncation process, it was reduced to $99926$.
 (b) Final histogram after the stochastic filling step for 1 realization of a network (filled green symbols). The same histogram in logarithmic bin (open black circles).  
 In both panels, the red straight line is a pure power-law with
 $\gamma \sim 3.45$,   $\langle k \rangle=6$.  The inset in (b) shows t the filling probability $r(k)$ and the red line that decays as $1/k^\gamma$ was drawn for comparison. 
 }
    \label{fig:trunc}
\end{figure}

 Figure \ref{fig:trunc}(a) shows the intermediate (non-normalized) distribution of degrees $p_0(k) \sim k^{-\gamma}$ after this deterministic process (hence a single realization is shown). For this network we used $N=10^5$, $\gamma \sim 3.45$ and $\langle k \rangle=6$, but  the number of vertices was reduced to $99926$.
  As $k$ increases, it becomes more evident that $p_0(k)\le p^*(k)$, returning values below the original (normalized) degree distribution. Notice that, for large enough $k$,  we can have  
  $\lfloor Np^*(k)\rfloor=0$, while $\lfloor Np^*(k)\rfloor \gg 1$ for small $k$.

The number of missing nodes is $N-\sum_k\lfloor Np^*(k) \rfloor$.  
The degree sequence  needs to be completed, to fulfill the desired  network size $N$ and  average degree $\langle k \rangle$.
The nonnegative fractional part $0 \le r(k) \equiv Np^*(k)-\lfloor N p^*(k) \rfloor \le 1 $ can be used as a filling probability, to draw the remaining elements of  the degree sequence, according to the following algorithm: 
\begin{enumerate}

\item If $\sum_k \lfloor N p^*(k) \rfloor < N$ and $\sum_k k < \langle k \rangle N$, a number $k$ is  uniformly drawn from the interval $[k_{min }, k_{max}]$, and we decide if it will be added to the sequence or not with probability $r(k)$.
Moreover, each value of $k$  can be selected only once, then if the drawn value has already been used, a new one is drawn.

   \item If after some iterations there is only one missing node, 
   then  $\langle k \rangle N -\sum_{k} k = k'\in [k_{min}, k_{max}]$.  
 In this case, the degree $k'$ is added to the sequence even if it has been drawn before. The only case in which a repetition is allowed.  If the last missing node $k'\notin [k_{min}, k_{max}]$, then we start it over from step 1.

\end{enumerate}

Figure \ref{fig:trunc}(b) presents the distribution of degrees $p(k)$, after filling the missing nodes generated in panel (a) for a single realization. The  average degree is exactly $\langle k \rangle=6.0$. 
In the example the degree distribution is a pure power-law 
but of course the procedure can be applied to any degree distribution.

It is interesting to note that the probability distribution  $r(k)$ to fill the missing values is nearly uniform up to the value of $k$ for which $\lfloor N p^* (k)\rfloor=0$, in which case $r(k)= N p^* (k)$, hence decaying as a power law for large $k$ (see inset). Let us mention that $r(k)$ has previously  been used to introduce large-degree sites (hubs) via the configuration model in power-law degree distributions~\cite{maria-BA-congigmodel}, although not to control the average degree. 
Notice that this procedure to eliminate fluctuations is not exclusive of power-law networks and in principle can be applied to any $p(k)$.

\section{Final remarks}
\label{sec:final}

We have analyzed two different modifications of the degree distribution that allow us to adjust the average degree $\langle k \rangle$, preserving the power-law character of the distribution. We considered two simple forms of modifying the low-$k$ region, but, of course, other shapes might also be used to produce similar results. 
Additionally, we have presented a procedure to eliminate fluctuations in the  average degree, which can be applied to any degree distribution.

Controlling $\langle k \rangle$  is important since correlations and other structural properties can be affected by the average connectivity~\cite{ramos2013random,serrano2005tuning}. 
It is particularly relevant when an artificial network is used as a substrate on top of which the dynamics of a complex system  evolves. 
Fixing the average degree in synthetic networks may be also useful for comparisons with real ones, overcoming the difficulties exposed in Ref.~\cite{brain2010}. A further advantage of applying the proposed schemes (either I or II) is that they help to eliminate fragments disconnected from the giant component, as soon as the low values of $k$ become less probable. 
The  procedures we proposed to tune the average degree  can be directly adapted to  adjust the value of  other finite moments of the degree distribution.

Let us finally mention that another way to built networks with a power-law degree distribution is using growth techniques. For example, the generalized Barabasi-Albert (GBA) model~\cite{BA-general} allows in principle to adjust the degree-distribution exponent $\gamma$ and average degree $\langle k \rangle$, for a given size of the network, however there are restrictions in the values that can be achieved (for instance, when $\gamma=3.5$,  $\langle k \rangle > 4$ is not possible in GBA).  
For instance in the particular case of a standard Barabasi-Albert network~\cite{BA}, 
with $\gamma=3$, only even values of $\langle k \rangle$ are obtained in the limit of large $N$.
Moreover, although GBA networks present a negligible Pearson coefficient~\cite{newman-assortative}, it is known that these graphs possess $k$-dependencies~\cite{BA_correlations,rewire_correlations}.

We have verified that the  modifications introduced in the degree distribution with structural cut-off do not introduce  $k$-dependencies in the nearest-neighbor degree $k_{nn}(k)$ and the  local clustering coefficient $c(k)$. Although choosing other form of the cutoff may introduce $k$-dependencies, the method to tune $\langle k \rangle$ is still effective.

We acknowledge partial financial support from Brazilian agencies CAPES (code 001), CNPq and Faperj.

%%%%%%%%%%%%%%%%%%%%%%%%%%%%%%%%%%%%%%

%%%%%%%%%%%%%%%%%%


\begin{thebibliography}{99}

\bibitem{watts}
Watts, D. J., 
{\it A simple model of global cascades on random networks}, Proc. Natl. Acad. Sci. USA 99, 5766–5771 (2002).
%\url{https://doi.org/10.1073/pnas.082090499}


\bibitem{extremists}
M. Ramos, J. Shao, S.D.S. Reis, C. Anteneodo, J.S. Andrade, S. Havlin, H.A. Makse, 
{\it How does public opinion become extreme?}, 
Sci. Rep. 5, 10032 (2005).
%\url{https://doi.org/10.1038/srep10032}

 
\bibitem{qvoter}
A.R. Vieira, A.F. Peralta, R. Toral, M. San Miguel, C. Anteneodo, 
{\it Pair approximation for the noisy threshold $q$-voter model}, 
Phys. Rev. E  101,  052131 (2020). %\url{10.1103/PhysRevE.101.052131}

 
\bibitem{silvio}
A. Mata, R.S. Ferreira, S. C. Ferreira,  
{\it Heterogeneous pair-approximation for the contact process on complex networks},
{New Journal of Physics} 16, 053006 (2014). 

\bibitem{cooperation}
C.-L. Tang, W.-X. Wang, X. Wu, and B.-H. Wang, 
{\it Effects of average degree on cooperation in networked evolutionary game}, Eur. Phys. J. B 53, 411–415 (2006) 
%DOI: 10.1140/epjb/e2006-00395-2



\bibitem{brito_correlations}
J. B. de Brito, C. I. N. Sampaio Filho,
A. A. Moreira, J. S. Andrade Jr.,
{\it Characterizing the intrinsic correlations of scale-free networks},
Int. J. Mod. Phys. C, 27, 3, 1650024 (2016).
%DOI: 10.1142/S0129183116500248
 
 
\bibitem{brain2010}
B.C.M. van Wijk1, C.J. Stam, A. Daffertshofer, 
{\it Comparing brain networks of different size and connectivity density using graph theory},
PLoS ONE 5(10) e13701 (2010).
% DOI https://doi.org/10.1371/journal.pone.0013701

\bibitem{CM} 
 M. Molloy, B. Reed, 
 {\it A critical point for random graphs with a given degree sequence}, 
 Random Structures \& Algorithms. 6 (2–3): 161–180 (1995).  % doi:10.1002/rsa.3240060204. I 
 
\bibitem{newman}
M. E. J. Newman,
{\it  Networks: an introduction}, 
(Oxford University Press, Oxford, 2018).


\bibitem{rewire_correlations}
J. Qu, S. J. Wang, M. Jusup, Z. Wang,
{\it Effects of random rewiring on the degree correlation of scale-free networks},
Sci. Rep. 5, 15450 (2015). 
%https://doi.org/10.1038/srep15450



\bibitem{mendes2002}
S. N. Dorogovtsev, and J. F. F. Mendes, 
{\it Evolution of networks}, Adv. Phys. 51, 1079 (2002).


\bibitem{UCM}   
M. Catanzaro, M. Bogu\~n\'a and R. Pastor-Satorras, 
{\it  Generation of uncorrelated random scale-free networks},
Phys. Rev. E 71, 027103 (2005).
%\url{https://doi.org/10.1103/PhysRevE.71.027103} 

 

\bibitem{UCM2004}
M. Bogu\~n\'a, R. Pastor-Satorras, A. Vespignani, 
{\it  Cut-offs and finite size effects in scale-free networks},
Eur. Phys. J. B 38, 205–209 (2004)
%\url{https://doi.org/10.1140/epjb/e2004-00038-8 }
 

 


 
 
\bibitem{newman-chapter}
M. E. J. Newman, in Handbook of Graphs and Networks:
From the Genome to the Internet, edited by S. Bornholdt and
H. G. Schuster  (Wiley-VCH, Berlin, 2003), pp. 35–68.
%\url{https://www.wiley.com/en-us/exportProduct/pdf/9783527606337}


\bibitem{serrano2007correlations}
M. A. Serrano, M. Bogu\~n{\'a}, R. Pastor-Satorras, A. Vespignani,
{\it Large scale structure and dynamics of complex networks:  From information technology to finance and natural sciences} (World Scientific Publishing Co. Ltd), pages 35--65 (2007). 
%https://doi.org/10.1142/6455
   
\bibitem{maria-BA-congigmodel}   
M. L. Bertotti and G. Modanese,
{\it The configuration model for Barabasi-Albert networks},
Applied Network Science 4, 32 (2019)
% https://doi.org/10.1007/s41109-019-0152-1

\bibitem{ramos2013random}
M. Ramos, C. Anteneodo,
{\it Random degree--degree correlated networks},
{Journal of Statistical Mechanics: Theory and Experiment} P02024 (2013).

\bibitem{serrano2005tuning}
M. A. Serrano  and M. Bogu\~n\'a, 
{\it Tuning clustering in random networks with arbitrary degree distributions}
Phys. Rev. E 72, 036133 (2005).

\bibitem{BA-general}
R. Albert and A.L. Barab\'asi,
{\it Topology of Evolving Networks: Local Events and Universality},
Phys. Rev. Lett. 85, 5234 (2000).
%DOI:https://doi.org/10.1103/PhysRevLett.85.5234

\bibitem{BA}
A.L. Barab\'asi,R. Albert ,
{\it  Emergence of scaling in random networks},  Science 286 (5439), 509 (1999).  %DOI: 10.1126/science.286.5439.509  
 

\bibitem{newman-assortative}
M. E. J. Newman,
{\it Assortative Mixing in Networks},
Phys. Rev. Lett. 89, 208701 (2002).
%DOI: https://doi.org/10.1103/PhysRevLett.89.208701

\bibitem{BA_correlations}
B. Fotouhi and M. G. Rabbat, 
{\it Degree correlation in scale-free graphs}, Eur. Phys. J. B 86, 510 (2013). 
% https://doi.org/10.1140/epjb/e2013-40920-6


\end{thebibliography}
\end{document}